\documentclass[reprint,twocolumn,superscriptaddress,showpacs,nofootinbib,notitlepage]{revtex4-1}

\usepackage{graphicx}
\usepackage{latexsym,amsmath,amssymb,lmodern,float,url}
\usepackage{natbib}
\usepackage{color}
\usepackage{microtype}

\usepackage[colorlinks=true,backref=false, linktocpage=true,
citecolor=blue,urlcolor=blue,linkcolor=blue,pdfpagemode=UseOutlines]{hyperref}

\hypersetup{%
  bookmarksnumbered=true,
  pdftitle = {},
  pdfsubject = {},
  pdfauthor = {},
  pdfkeywords = {}
}

\def\Tr {\operatorname{{Tr}}}

\renewcommand{\d}{\mathrm d}

\begin{document}
\title{Perturbative Removal of a Sign Problem}

\author{Scott Lawrence}
\email{scott.lawrence-1@colorado.edu}
\affiliation{Department of Physics, University of Colorado, Boulder, CO 80309, USA}
\date{\today}

\begin{abstract}
This paper presents a method for alleviating sign problems in lattice path integrals, including those associated with finite fermion density in relativistic systems. The method makes use of information gained from some systematic expansion --- such as perturbation theory --- in order to accelerate the Monte Carlo. The method is exact, in the sense that no approximation to the lattice path integral is introduced. Thanks to the underlying systematic expansion, the method is systematically improvable, so that an arbitrary reduction in the sign problem can in principle be obtained. The Thirring model (in $0+1$ and $1+1$ dimensions) is used to demonstrate the ability of this method to reduce the finite-density sign problem.
\end{abstract}

\maketitle

\section{Introduction}
Lattice Monte Carlo methods are able to provide nonperturbative access to observables in quantum field theories. They are unique in this respect for many strongly coupled theories. Under certain circumstances, such as at finite density of relativistic fermions and the Hubbard model away from half-filling, lattice methods are made dramatically less efficient by the so-called sign problem. This sign problem is a central obstacle to first-principles calculations in many regimes of strongly coupled theories, including \emph{ab initio} studies of the nuclear equation of state.

In lattice field theory, spacetime is treated as discrete and observables are obtained from the high-dimensional lattice path integral. Lattice field theory is ordinarily used to study a system in thermal equilibrium, and the partition function is written as $Z = \int \mathcal \mathcal D A \;e^{-S(\phi)}$, where $S$ is the (Euclidean) action and the integral is taken over all configurations of a field $A$. Observables are given by various derivatives of the logarithm of the partition function. These derivatives are ordinarily sampled by importance sampling, which hinges on the treatment of the normalized Boltzmann factor $e^{-S}/Z$ as a probability distribution. For some systems, including those with a finite density of relativistic fermions, the action $S$ is complex, and this is not possible --- this is the sign problem.

Importance sampling commonly takes a polynomial amount of time in the spacetime volume being simulated (although this is proven only in a few cases~\cite{jerrum1993polynomial,guo2016random,Collevecchio_2016}). Importance sampling can be modified to work even where $S$ is complex, but at the cost of efficiency. In this modification, the ``quenched" Boltzmann factor $|e^{-S}| / Z$ is treated as a probability with respect to which sampling is performed. Ordinary expectation values are obtained in terms of quenched expectation values: $\langle\mathcal O\rangle = \langle\mathcal O e^{-i S_I}\rangle_Q/\langle e^{-i S_I}\rangle_Q$. The loss of efficiency comes primarily from the denominator. The average of the exponential of the imaginary part of the action, often termed the ``average phase", is equal to the ratio of the physical to quenched partition functions $Z/Z_Q$, and characteristically scales like $e^{-\beta V}$. Resolving this exponentially small quantity, by averaging many quantities of unit magnitude, requires $\sim e^{2\beta V}$ samples; thus the reweighting procedure incurs an exponential cost in the volume. This failure affects a wide variety of models, and a correspondingly wide variety of methods have been proposed to mitigate it: complex Langevin~\cite{Aarts:2008rr}, the density of states method~\cite{Langfeld:2016mct}, canonical methods~\cite{Alexandru:2005ix,deForcrand:2006ec}, reweighting methods~\cite{Fodor:2001au}, series expansions in the chemical potential~\cite{Allton:2002zi}, fermion bags~\cite{Chandrasekharan:2013rpa}, field complexification~\cite{Alexandru:2020wrj}, and analytic continuation from imaginary chemical potentials~\cite{deForcrand:2006pv}.

In this paper we will examine a new method, inspired by two observations: first, that the partition function is unchanged if a function that integrates to zero is added to the Boltzmann factor, and second, that lattice methods can encounter a fatal sign problem even in regimes under good control by perturbation theory (or any other systematic expansion). To any fixed order in perturbation theory, the sign problem can be (non-uniquely) identified with some oscillating part of the Bolzmann factor which integrates to zero, and this part can then be subtracted off, without changing the partition function or any observables. In fact, we will see that this subtraction can be performed in such a way that even the nonperturbative partition function and observables remain unchanged. Where the model is under good control by perturbation theory, meaning that the partition function is well-approximated by the integral of a perturbative expansion of the integrand, this subtraction is nearly the entire sign problem. In regimes where perturbation theory is a poor approximation, we may hope to isolate and remove a single component of the sign problem, thereby improving the efficiency of the necessary nonperturbative calculation.

The method described in this paper exhibits two favorable characteristics worth noting before we begin. Firstly, it is an \emph{exact} method, in the sense that the modified form of the partition function is precisely equal to the original, physical form. As a consequence, all observables retain their physical values, and the only errors are statistical ones associated to the sampling process. This is true regardless of the quality of the systematic expansion used: the removal of the sign problem is approximate, but the observables computed are exact. Secondly, although the removal of the sign problem is approximate, it is \emph{systematically improvable}. If a certain order in perturbation theory does not yield a sufficiently moderate sign problem, a higher order can in principle be used. As long as the expansion converges (on the lattice), a sufficiently high order is guaranteed to remove the sign problem to any desired degree. Of course, an exponential cost is associated with going to higher orders in most expansions, and it is to be expected that this property of systematic improvability is not a practical way to solve many problems, as it merely trades one exponential cost for another. Nevertheless, this is an unusual and promising combination.

This paper uses the Thirring model~\cite{thirring1958soluble} in $0+1$ and $1+1$ dimensions as a testbed for the method of subtractions. This model has frequently been used, in varying dimensions, to test methods for treating the fermion sign problem in the past, including complexification~\cite{Alexandru:2016ejd,Alexandru:2015sua} and complex Langevin~\cite{Pawlowski:2013gag}.

In the next section, the general method of subtractions is described in detail, with an emphasis on subtractions that are constructed via some systematic expansion. In Sec.~\ref{sec:mechanics}, the heavy-dense limit is used to construct a subtraction for the Thirring model in $0+1$ dimensions. This is extended in Sec.~\ref{sec:fields}, where the $1+1$-dimensional Thirring model is treated with a variety of expansions. A nonperturbative method of optimizing subtractions is described in Sec.~\ref{sec:opt}. Finally we conclude in Sec.~\ref{sec:discussion}, discussing in particular a relation between this method and the method of field complexification.

\section{General Method}\label{sec:general}
For brevity, let us write the Boltzmann factor as $f(A) \equiv e^{-S(A)}$, so that the unmodified form of the partition function is $Z = \int \mathcal D A\; f(A)$. If we let $g(A)$ be some function which integrates to $0$ (e.g.\ a total derivative of a function with appropriate behavior on the boundary of configuration space), then the numerical value of the partition function is unmodified by the subtraction of $g(A)$ from the Boltzmann factor:
\begin{equation}
Z = \int \mathcal D A\; f(A)
=  \int \mathcal D A\; f(A) - g(A)
\text.
\end{equation}
The quenched partition function, and therefore the average phase $\langle \sigma\rangle \equiv Z/Z_Q$, is generically changed by this operation. Therefore, a suitable $g(A)$ may improve the sign problem. In fact, a subtraction always exists which removes the sign problem entirely:
\begin{equation}\label{eq:perfect}
g_{\mathrm{ideal}}(A) = f(A) - \frac{\int \mathcal D A' \; f(A')}{\int \mathcal D A'}
\text.
\end{equation}
This particular subtraction is unusable in practice, as computing it requires exact knowledge of the partition function. Indeed, using this subtraction is equivalent to performing the entire computation analytically.

Particularly in the case where $g$ is constructed from a perturbative expansion (described below) this method can be thought of as splitting the path integrand into a few terms, and integrating some analytically. In the case of the ideal subtraction of Eq.~(\ref{eq:perfect}), the entire path integral is performed analytically.

Once a subtraction is selected, it remains to compute an observable. We must express $\langle \mathcal O\rangle$ (an expectation value over $f$) as an expectation value taken over the distribution $f-g$. It is tempting to write
\begin{equation}
\langle\mathcal O\rangle
=
\frac{\int \mathcal D A\; (f(A) - g(A)) \frac{\mathcal O(A) f(A)}{f(A) - g(A)}}{\int \mathcal D A\; f(A) - g(A)}
\text.
\end{equation}
This equation is correct, but not useful for computing the expectation value, as the measurement of the modified observable encounters a signal-to-noise problem comparable to the original sign problem. This is particularly clear in the case of $\mathcal O = 1$, where the numerator is equal to $\int \mathcal D A\; f(A)$, the highly oscillatory integral we wanted to avoid in the first place.

Consider a conjugate variable $\xi$ to $\mathcal O$, such that $\langle \mathcal O\rangle = \frac{\partial}{\partial \xi}\log Z$. The previous approach corresponds to treating $g$ as constant in $\xi$. Instead, take $g$ to vary with $\xi$, in such a way that $\int g = 0$ for any value of $\xi$. The desired expectation value is now
\begin{equation}\label{eq:obs}
\langle \mathcal O\rangle
=
\frac{\int \mathcal D A\; \mathcal O(A) f(A) - \frac{\partial}{\partial \xi}g(A)}{\int \mathcal D A\; f(A) - g(A)}
\text,
\end{equation}
which does not necessarily (and does not in practice, as we will see) suffer from the same magnitude of signal-to-noise problem.

We now discuss how to construct a suitable subtraction $g(A)$ in a systematic manner. One strategy is to attempt to approximate Eq.~(\ref{eq:perfect}) as closely as possible, with an analytic expansion. For the purposes of removing the sign problem, however, it is sufficient to replace $f(\cdot)$ in Eq.~(\ref{eq:perfect}) by just the part of the Boltzmann factor that oscillates. Removing the oscillations will cure the sign problem, even if the rest of the partition function is not approximated well at all.

To make this concrete, suppose a perturbative expansion of $f(A)$
\begin{equation}
f(A) = f_0(A) + \lambda f_1(A) + \frac{\lambda^2}{2} f_2(A) + \cdots
\end{equation}
is available, such that the partition functions at low order are readily (perhaps analytically) obtained. Defining $Z_n = \int \mathcal D A\; f_n(A)$, we can construct a wide variety of functions which integrate to $0$ and approximate various parts of the original Boltzmann factor. It is often convenient to pick (some linear combination of)
\begin{equation}\label{eq:perturbative-subtraction}
g_n(A) = f_n(A) - \frac{f_0(A)}{Z_0}Z_n
\text.
\end{equation}
The factor of the free theory Boltzmann factor is somewhat arbitrary --- any function of $A$ with unit integral will do.

This procedure does not depend on the precise nature of the systematic expansion. Our first application of this method (in Sec.~\ref{sec:mechanics}) will use the heavy-dense limit to construct a subtraction, instead of an expansion around free field theory.

Because the subtracton was constructed from a systematic expansion, $g_n$ naturally depends on $\xi$. Applying Eq.~(\ref{eq:obs}) to this construction, the physical expectation value of $\mathcal O$ is given by
\begin{equation}\label{eq:obs-perturbative}
\langle \mathcal O \rangle
=
\left\langle \frac{\mathcal O f - \frac{\partial}{\partial \xi}f_n + \frac{f_0}{Z_0}\frac{\partial}{\partial \xi}Z_n}{f-g_n}\right\rangle_{\!f-g_n}
\text.
\end{equation}
Note that it is not in general true that $\frac{\partial}{\partial \xi} f_n = \mathcal O f_n$, nor is it generally true that the same derivative of $\log Z_n$ yields a perturbative expectation value.

In deriving this expression, we have chosen for convenience not to let $f_0$, and therefore $Z_0$, vary with $\xi$. This, like the precise manner of constructing the subtraction, is an arbitrary choice. We will not, in this paper, explore the question of what the optimal construction of a modified observable is.

Of course, even after the subtraction, a residual sign problem typically remains, which is addressed by reweighting.

\section{Quantum Mechanics}\label{sec:mechanics}
\begin{figure*}
\begin{minipage}{0.31\textwidth}
\centering
\includegraphics{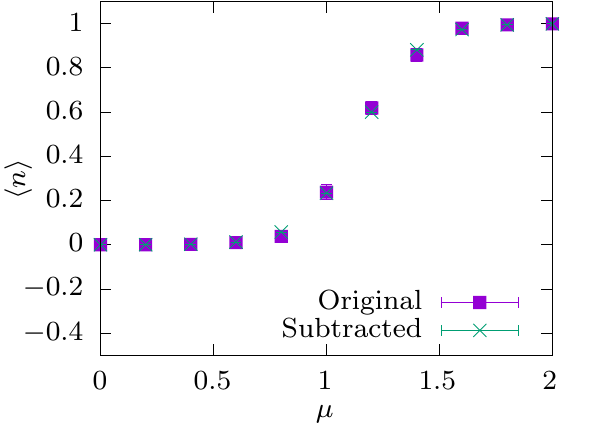}
\end{minipage}
\begin{minipage}{0.31\textwidth}
\centering\includegraphics{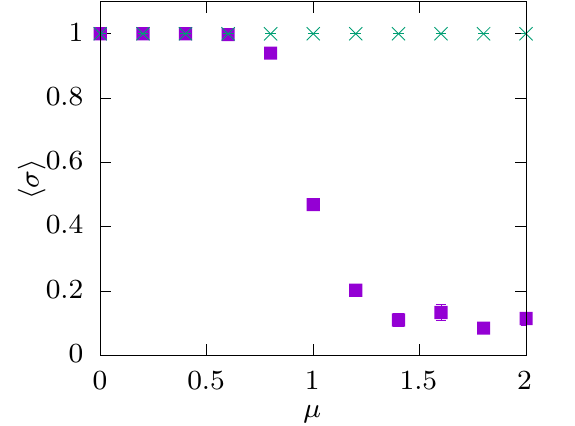}
\end{minipage}
\begin{minipage}{0.31\textwidth}
\centering\includegraphics{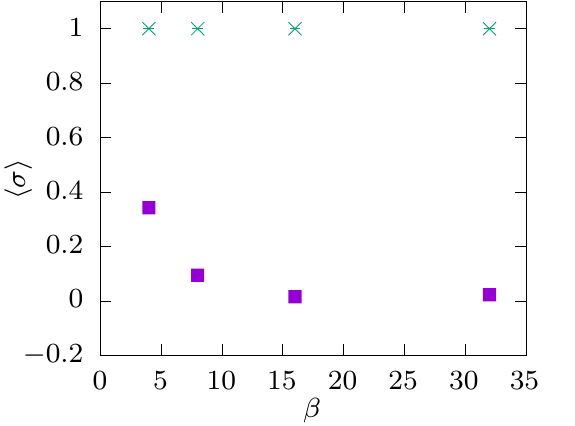}
\end{minipage}
\caption{\label{fig:mechanics} The subtraction method as applied to the $0+1$-dimensional Thirring model. The leftmost plot shows the density as a functon of chemical potential with $\beta = 8$, $m=1$, and $g^2 = 0.2$. The exact result is from~\cite{Pawlowski_2015}. The center plot shows the average phase, again as a function of $\mu$, for the same parameters. On the right is the average phase for $\mu=1.8$ as a function of inverse temperature $\beta$.}
\end{figure*}

In this section we demonstrate the method on a $0+1$-dimensional variant of the Thirring model. Described in~\cite{Alexandru:2015xva,Alexandru:2015sua}, this model is defined by the lattice action
\begin{equation}\label{eq:mechanics-action}
S = \frac {1}{2 g^2} \sum_t \left(1 - \cos A(t)\right) - \log \det K[A].
\end{equation}
The Dirac matrix $K[A]$ is given by
\begin{widetext}
\begin{equation}
K[A]_{t t'} = \frac 1 2
\big[
e^{\mu + i A(t)} \delta_{(t+1)t'}
- e^{-\mu-iA(t')} \delta_{(t'+1)t}
- e^{\mu + i A(t)} \delta_{tN}\delta_{t'1}
+ e^{-\mu-iA(t')} \delta_{t1}\delta_{t'N}
\big] + m \delta_{tt'}
\text.
\end{equation}
\end{widetext}
Above, $m$ is the bare mass and $g$ a coupling constant; we are implicitly working in units where the lattice spacing is $1$, so that the number of sites is equal to the inverse temperature $\beta$. The sign problem, created by the chemical potential $\mu$, is portrayed in Fig.~\ref{fig:mechanics}; the average phase decays exponentially with the inverse temperature, and so the cost of calculations increases exponentially with the same.

A suitable subtraction is provided by the heavy-dense limit of $\mu\rightarrow\infty$. The Dirac matrix can be expanded via the polymer representation~\cite{montvay1997quantum}, and the dominant term of $\det K$ in the limit of large $\mu$ is
\begin{equation}
\det K = e^{\beta\mu} (2^{-\beta} e^{i \sum_t A(t)} + O(e^{-\beta \mu}))
\text.
\end{equation}
We will use the leading-order term as our subtraction:
\begin{equation}\label{eq:f1qm}
f_1(A) = e^{\frac 1 {2 g^2} \sum_t \cos A(t)} \times 2^{-\beta} e^{\beta\mu + i \sum_t A(t)}
\text.
\end{equation}
Integrating over all fields yields the leading-order partition function
\begin{equation}\label{eq:Z1qm}
Z_1 = e^{\beta\mu} \left[\pi I_1\left(1 / 2 g^2\right)\right]^\beta\text.
\end{equation}
(Here and throughout, $I_\nu(\cdot)$ denotes the modified Bessel function of the first kind, of order $\nu$.)

For the scaling factor $f_0(A) / Z_0$ in Eq.~(\ref{eq:perturbative-subtraction}) we could simply choose $(2\pi)^{-\beta}$, but it is convenient in this case to use the bosonic part of the Boltzmann factor:
\begin{equation}\label{eq:scaling}
\frac{f_0(A)}{Z_0}
=
\frac{\exp\left(\frac 1 {2 g^2} \sum_t (1 - \cos A(t))\right)}{\left[2 \pi I_0(1/2g^2)\right]^\beta}
\text.
\end{equation}

The observable we will focus on is the number density, defined as $\langle n \rangle = \beta^{-1} \frac{\partial}{\partial \mu}\log Z$. In order to measure this observable with the subtraction method, we need the $\mu$-derivatives of $f_1$ and $Z_1$ as per Eq.~(\ref{eq:obs-perturbative}). Happily, in this case they are particularly simple: $\frac{\partial}{\partial\mu} f_1 = f_1$ and $\frac{\partial}{\partial\mu} Z_1 = Z_1$. This reflects the fact that, in the heavy-dense limit, the density is $1$ regardless of temperature.

To summarize, before performing the subtraction, the partition function was written $Z = \int e^{-S}$ with the action $S$ defined by Eq.~(\ref{eq:mechanics-action}). The modified form of the partition function is
\begin{widetext}
\begin{equation}
Z = \int \mathcal D A \;
\underbrace{\exp\left(\frac 1 {2 g^2} \sum_t \left(1 - \cos A(t)\right)\right)}_{f_0}
\bigg[
\underbrace{\det K}_{f / f_0}
- \underbrace{2^{-\beta} e^{\beta \mu + i \sum_t A(t)}}_{f_1 / f_0}
+ \underbrace{\frac{e^{\beta\mu} \left[\pi I_1(1/2g^2)\right]^\beta}{\left[2 \pi I_0(1/2g^2)\right]^\beta}}_{Z_1 / Z_0}
\bigg]
\text,
\end{equation}
\end{widetext}
where the scaling factor $f_0$ and its integral $Z_0$ are defined by Eq.~(\ref{eq:scaling}), and the subtraction is constructed from the heavy-dense term $f_1$ and its integral $Z_1$, given in Eqs.~(\ref{eq:f1qm}) and (\ref{eq:Z1qm}).

While numerically identical, this form is hoped to have a reduced sign problem. The density is given by the expectation value, taken in the subtracted ensemble,
\begin{equation}
\beta \langle n \rangle = \left<
\frac{
\Tr K^{-1} \frac{\partial K}{\partial \mu} - f_1 + f_0 \frac{Z_1}{Z_0} 
}{
f - f_1 + f_0 \frac{Z_1}{Z_0}
}
\right>_{\!f-g}
\text.
\end{equation}

The results of this procedure are shown in Fig.~\ref{fig:mechanics}. Specially in $0+1$ dimensions, the sign problem is no longer exponential in the volume, but rather improves slightly as $\beta$ is increased. This is not to be expected to hold true for higher dimensional theories. In general, the exponential difficulty of the sign problem will not be removed by the subtraction method, but merely ameliorated. (In the case of the particular model at hand, it is possible to construct a subtraction that entirely removes the sign problem, but only because the entire partition function is analytically known.)

Lastly, note that all data points in Fig.~\ref{fig:mechanics} are constructed from $10^3$ samples. The data points calculated with the subtraction have much smaller error bars (for $\mu=2.0$, the error bar width is $\sim 10^{-14}$) even than the sign-free $\mu=0$ data point without the subtraction; this procedure has improved the signal-to-noise ratio in addition to reducing the sign problem. In the limit of the ideal subtraction of Eq.~\ref{eq:perfect}, there is no variance remaining in the observable, and a single measurement yields the exact answer.

\section{Field Theory}\label{sec:fields}
\begin{figure*}
\centerline{
\begin{minipage}{0.45\textwidth}
\centering\includegraphics{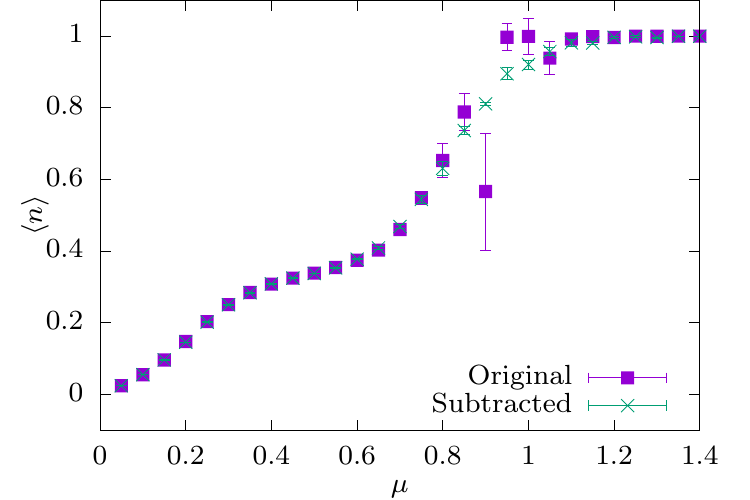}
\end{minipage}
\begin{minipage}{0.45\textwidth}
\centering\includegraphics{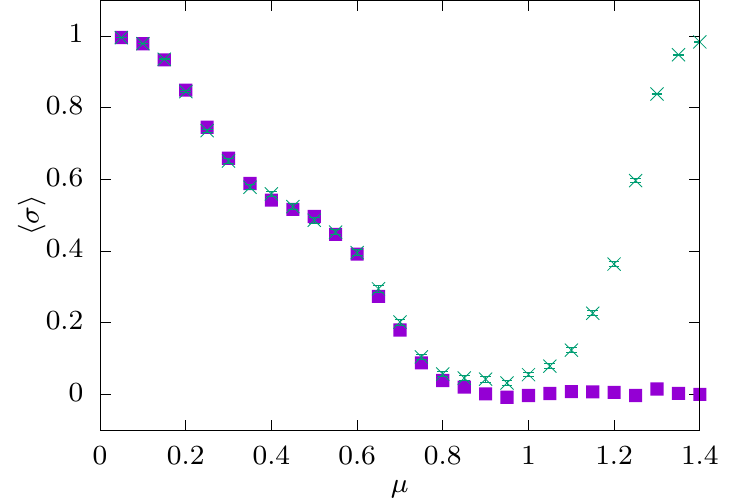}
\end{minipage}
}
\caption{Simulation of the $1+1$-dimensional Thirring model on a $12\times 6$ lattice with $m = 0.15$, $g^2=0.3$. The left plot shows the density as a function of chemical potential, and on the right are the corresponding average phases. Each data point is backed by $10^4$ samples.\label{fig:hd}}
\end{figure*}

We now move to the $1+1$-dimensional Thirring model with staggered fermions. The lattice action of this model is~\cite{Alexandru:2016ejd}
\begin{equation}
S = \sum_{x,\nu=0,1} \frac{2}{g^2} (1 - \cos A_\nu(x)) - \log \det K[A]
\end{equation}
with the Dirac matrix now defined by
\begin{align}
K[A]_{xy} = m \delta_{xy} + \frac 1 2 \sum_{\nu=0,1} & \eta_\nu e^{i A_\nu(x) + \mu \delta_{\nu,0}} \delta_{x+\nu,y}
\\\nonumber
- &\eta_\nu e^{-i A_\nu(y) - \mu \delta_{\nu,0}} \delta_{y+\nu,x}
\text,
\end{align}
where as before $m$ is the bare mass, $g$ the coupling, and $\mu$ the chemical potential. The staggered fermions are defined by $\eta_0 = (-1)^{\delta_{0 x_0}}$ and $\eta_1 = (-1)^{x_0}$. As in the $0+1$-dimensional model, a sign problem is created at $\mu \ne 0$.

The first subtraction procedes from the same heavy-dense limit we used for the quantum mechanical model above. As before, we define $f_0 = e^{ \frac{2}{g^2} \sum_{x,\nu} \cos A_\nu(x)}$. The leading-order term in the heavy-dense expansion is
\begin{equation}
f_1 = e^{\frac{2}{g^2} \sum_{x,\nu}\cos A_\nu(x)}
2^{-\beta L} e^{\beta L \mu + i \sum_x A_0(x)}
\end{equation}
which, when integrated over all fields, yields the partial partition function
\begin{equation}
Z_1 = e^{\beta L \mu} 2^{-\beta L} \left(2 \pi I_0(2/ g^2) I_1(2/ {g^2})\right)^{\beta L}
\text.
\end{equation}
At this order in the heavy-dense expansion, everything takes the form of $L$ copies of the quantum mechanical model above. In particular, the $\mu$-derivatives of $f_1$ and $Z_1$ are $L f_1$ and $L Z_1$, respectively.

The results of simulating with the leading-order heavy-dense subtraction, on a $12\times 6$ lattice, are shown in Fig.~\ref{fig:hd}. Without the subtraction, the sign problem falls to be indistinguishable from $0$ ($\lesssim 10^{-2}$) by $\mu \approx 0.8$; after the subtraction, the sign problem is manageable from $\mu = 0$ through lattice saturation.

At the next order in the heavy-dense limit, the number of diagrams in the polymer representation is exponential in $\beta$. Therefore, it is not practical (barring another way of computing the NLO heavy-dense partition function) to use this expansion at higher orders. Another expansion to consider is the hopping expansion. However this expansion is also not practical for the purpose of removing a sign problem, as the lowest-order term in the hopping expansion that has a sign problem is at order $\kappa^\beta$.

\begin{figure}
\centerline{
\includegraphics{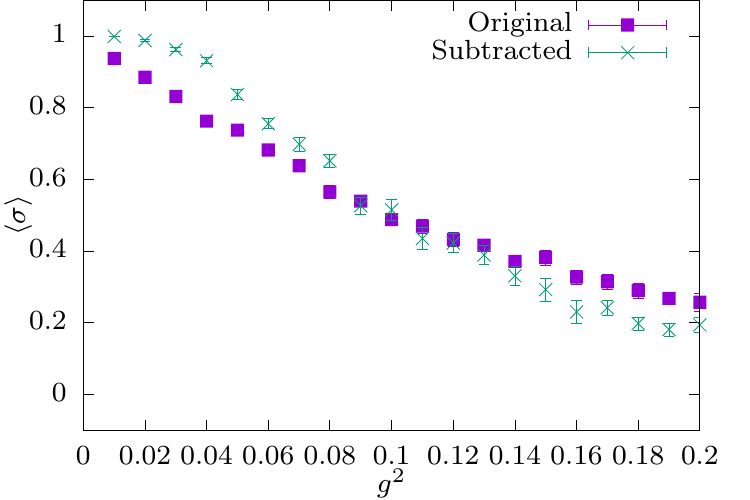}
}
\caption{The sign problem on a $6\times 6$ lattice with bare mass $m=0.15$ and chemical potential $\mu=1$, with and without the subtraction of Eq.~(\ref{eq:weaksub}).\label{fig:weak}}
\end{figure}

At small $g$, the auxiliary field is pegged to $A \sim 0$ by the $\cos A$ term in the action. As a result, it is possible to construct a ``weak-coupling'' expansion for the lattice Thirring model described here by Taylor expanding $\det K[A]$ in the fields $A$. The term first-order in $A$ makes a particularly convenient subtraction: as it is odd in $A$, it integrates to $0$, and the corresponding partial partition function $Z_1$ vanishes. The subtracted integrand of the partition function is
\begin{equation}\label{eq:weaksub}
f-g = f_0
\left[
\det K
- \det K_0 \Tr K_0^{-1} \left(\frac{\partial K}{\partial A}\right)_{A=0} A
\right]
\end{equation}
where $K_0$ is $K$ evaluated at $A=0$, and $f_0 = e^{\frac 2 {g^2} \sum_{x,\nu} \cos A_\nu(x)}$ as usual. Fig.~\ref{fig:weak} shows the magnitude of the sign problem on a $6\times 6$ lattice, as a function of the squared coupling constant, with and without this subtraction. A systematic improvement is visible at small values of the coupling; at sufficiently large value of $g^2$, the subtraction is no longer guaranteed to help.

\section{Nonperturbative Optimization}\label{sec:opt}
So far, we have described how a suitable subtraction can be engineered with the aid of a systematic expansion, such as the weak coupling or heavy-dense limit. Subtractions constructed in this manner need not be optimal, and it may be profitable to consider other possibilities. In this section we will see that it is possible to efficiently perform a nonperturbative optimization on a family of ansatz subtractions to find the one with the largest average phase. The method discussed here was used in a very similar form for optimizing manifolds of integration~\cite{Alexandru:2018fqp}, and has been applied (in one form or another) to several different field theories~\cite{Alexandru:2018ddf,Kashiwa:2019lkv,Ohnishi:2019ljc}.

Suppose we have a continuous family of actions $S_\alpha$ (the parameter $\alpha$ may have many components), such that the partition function $Z = \int e^{-S_\alpha}$ does not depend on $\alpha$. This is exactly the case if $\alpha$ defines a subtraction, or as in \cite{Alexandru:2018fqp}, a manifold of integration. Although the partition function has no dependence on $\alpha$, the quenched partition function and therefore the sign problem may. In general, computing the sign problem for any fixed $\alpha$ is computationally expensive. We would like to invest computational resources efficiently, performing a simulation with the value of $\alpha$ that has the mildest sign problem. However, finding such a value appears hard: it certainly isn't feasible to do a grid search, resolving the sign problem for each value of $\alpha$, in order to find the best one.

Consider performing gradient ascent on the logarithm of the average phase. Arbitrarily picking some initial $\alpha$, we would like to calculate $\frac{\partial}{\partial \alpha} \log \frac{Z}{Z_Q(\alpha)}$, which specifies the direction in which we should move. If we were to calculate this by finite differencing, we would need to resolve the sign problem at both $\alpha$ and $\alpha+\epsilon$, an expensive proposition. However, observe that
\begin{equation}
\frac{\partial}{\partial\alpha} \log \frac{Z}{Z_Q(\alpha)}
= - \frac{\partial}{\partial\alpha} \log Z_Q(\alpha)
\end{equation}
has the form of a derivative of the logarithm of the \emph{quenched} partition function, and the contribution of the physical $Z$ cancels entirely. The direction which most quickly alleviates the sign problem is a quenched expectation value, which can be computed without encountering a sign problem.

With this observation in hand, we see that it is possible to begin with a family of subtractions $g_\alpha$, and perform an efficient, sign-free gradient descent to find the optimal subtraction in that family. At this point, a (comparatively expensive) Monte Carlo can be performed, with high statistics to counter the remaining sign problem.

\begin{figure}
\centering
\includegraphics{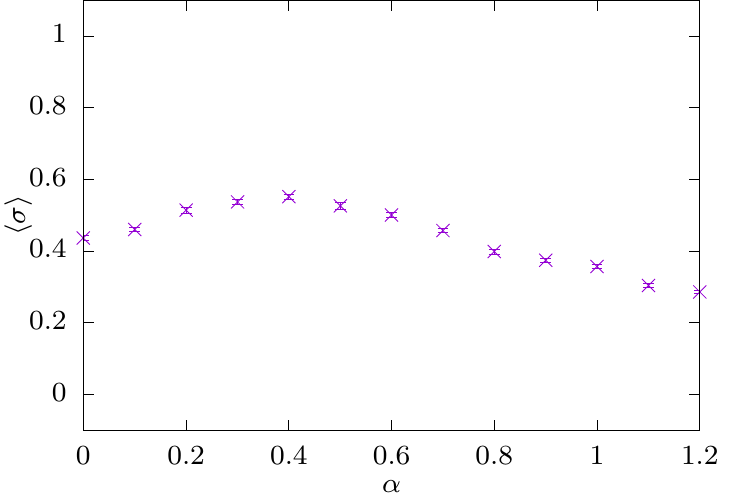}
\caption{The magnitude of the sign problem for a $4 \times 4$ lattice with $m=0.15$, $g^2 = 0.3$, and $\mu = 1$, as a function of the subtraction coefficient $\alpha$, using the subtraction of Eq.~(\ref{eq:weaksub-tunable}).\label{fig:subscan}}
\end{figure}

One motivation for this method stems from the ``weak-coupling'' subtraction of the previous method. The subtraction $g = f_1 - Z_1$ defined by Eq.~(\ref{eq:weaksub}) can be multiplied by an arbitrary coefficient $\alpha$, so that the integrand of the partition function is modified by
\begin{equation}\label{eq:weaksub-tunable}
- g_\alpha = - \alpha f_0
\det K_0 \Tr K_0^{-1} \left(\frac{\partial K}{\partial A}\right)_{A=0} A
\text.
\end{equation}
In the previous section, the coefficient used was implicitly $1$; as shown in Fig.~\ref{fig:subscan}, it turns out that this is not the optimal coefficient. The optimization procedure described above can be used to optimize this coefficient at scale. Note that for the example shown here, the full-magnitude first-order subtraction makes the sign problem \emph{worse} at $g^2=0.3$. However, nonperturbative optimization can reverse this, making the first-order subtraction useful even at this relatively large coupling.

\section{Discussion}\label{sec:discussion}

The method of subtractions described in this paper allows practical mitigation of sign problems associated to finite fermion density and real-time observables. The method is exact in the sense that it makes no additional approximations in the partition function. Furthermore, the removal of the sign problem, although only approximate, is systematically improvable.

This method is not unrelated to prior work. In particular, the method of field complexification~\cite{Alexandru:2020wrj} may be seen as a specific strategy for constructing a subtraction\footnote{In fact, the subtraction method was initially inspired by an attempt to extend the method of field complexification to the case of path integrals with discrete domains of integration.}. In that method, the original domain of the path integral --- $\mathbb R^N$, say --- is expanded to a complex space of twice the (real) dimension. In this case, the expanded space would be $\mathbb C^N$. By Cauchy's integral theorem, the path integral can now be performed over any $N$-real-dimensional manifold $\mathcal M \subset \mathbb C^N$ obtained by a smooth deformation from $\mathbb R^N$ (and with mild constraints at infinity, when the complex space is unbounded). Typically the new manifold is parameterized by the real plane via a function $\tilde\phi$ mapping field configurations $\phi\in\mathbb R^N$ to field configurations on $\mathcal M$, so that the deformed path integral is written
\begin{equation}
Z
= \int_{\mathbb R^N} \!\mathcal D \phi\; e^{-S[\phi]}
= \int_{\mathbb R^N} \!\mathcal D \phi\; e^{-S[\tilde\phi(\phi)]} \det \frac{\partial\tilde\phi}{\partial\phi}
\text.
\end{equation}
The difference between the two integrands is zero, and so can be viewed as a subtraction. Of course, in this view, every modification to the path integral that leaves the integration domain unchanged is a special case of the subtraction method.

We can also go a step further and note that the difference between the two integrands is a total derivative. Concretely, in one dimension, the difference between the two Boltzmann factors is
\begin{equation}
e^{-S[\tilde\phi(\phi)]} \det \frac{\partial\tilde\phi}{\partial\phi} - e^{-S[\phi]}
= \frac{\partial}{\partial \phi} \int_{\phi}^{\tilde\phi(\phi)} e^{-S[\phi']} \;\d\phi'
\text.
\end{equation}

It is notable that a well-chosen subtraction can resolve a sign problem even in cases where no manifold can. A simple example of a sign problem unremovable by any choice of manifold is the one-dimensional integral (which is to be considered a mock partition function)
\begin{equation}
Z = \int_{-\pi}^\pi \!\d\theta \; \left[ \cos(\theta) + \epsilon\right]
\text.
\end{equation}
The sign problem associated to this partition function becomes arbitrarily bad as $\epsilon$ is taken towards $0$. This sign problem was shown in \cite{Lawrence:2020irw} to be unremovable by any choice of integration contour. In fact, the original integration domain $S^1$ has a more mild sign problem than any other choice of domain. In this case, it's particularly easy to see that a subtraction of $\cos\theta$ completely removes the sign problem, where no manifold can. Thus the method of subtractions is strictly more powerful than that of complexification.

The manifold used in~\cite{Alexandru:2018fqp,Alexandru:2018ddf} to improve the sign problem of the Thirring model in $1+1$ and $2+1$ dimensions was motivated (post-hoc) by the leading-order term in the heavy-dense expansion. In~\cite{Lawrence:2018mve} it was shown that a manifold of that form can entirely remove the sign problem coming from that leading-order term. This choice of manifold is therefore equivalent to a subtraction constructed from that term.

The complexification method has been applied to real-time observables through the lattice Schwinger-Keldysh formalism~\cite{Alexandru:2016gsd}. The determination of real-time observables on the lattice remains a largely unexplored area. Future work should be able to apply the subtraction method to real-time calculations through the same formalism.

The success of the method described in this paper depends on the availabilty of a systematic expansion in which the sign problem can be seen. We have seen that several options exist for the Thirring model. Examining and making use of such expansions in other models is a critical next step.

We noted in Sec.~\ref{sec:mechanics} that in addition to improving the sign problem, the signal-to-noise ratio associated with the modified observable was improved from the one associated with the original observable. This was not explored further in this paper, but it suggests that the same or a similar method could be deployed explicitly for treating expensive signal-to-noise problems. It is not entirely surprising that this should be possible, as the closely related complexification method has recently been applied to noisy observables in Abelian gauge theory and complex scalar field theory~\cite{Detmold:2020ncp}.

\begin{acknowledgments}
I am indebted to Andrei Alexandru, Paulo Bedaque, and Henry Lamm for many useful conversations regarding the sign problem. I am also grateful to Henry Lamm for comments on an earlier version of this manuscript. This work was supported by the U.S.\ Department of Energy under Contract No.\ DE-FG02-93ER-40762, and subsequently by the U.S.\ Department of Energy under Contract No.\ DE-SC0017905.
\end{acknowledgments}
\bibliographystyle{apsrev4-1}
\bibliography{Self,References}

\begin{thebibliography}{27}%
\makeatletter
\providecommand \@ifxundefined [1]{%
 \@ifx{#1\undefined}
}%
\providecommand \@ifnum [1]{%
 \ifnum #1\expandafter \@firstoftwo
 \else \expandafter \@secondoftwo
 \fi
}%
\providecommand \@ifx [1]{%
 \ifx #1\expandafter \@firstoftwo
 \else \expandafter \@secondoftwo
 \fi
}%
\providecommand \natexlab [1]{#1}%
\providecommand \enquote  [1]{``#1''}%
\providecommand \bibnamefont  [1]{#1}%
\providecommand \bibfnamefont [1]{#1}%
\providecommand \citenamefont [1]{#1}%
\providecommand \href@noop [0]{\@secondoftwo}%
\providecommand \href [0]{\begingroup \@sanitize@url \@href}%
\providecommand \@href[1]{\@@startlink{#1}\@@href}%
\providecommand \@@href[1]{\endgroup#1\@@endlink}%
\providecommand \@sanitize@url [0]{\catcode `\\12\catcode `\$12\catcode
  `\&12\catcode `\#12\catcode `\^12\catcode `\_12\catcode `\%12\relax}%
\providecommand \@@startlink[1]{}%
\providecommand \@@endlink[0]{}%
\providecommand \url  [0]{\begingroup\@sanitize@url \@url }%
\providecommand \@url [1]{\endgroup\@href {#1}{\urlprefix }}%
\providecommand \urlprefix  [0]{URL }%
\providecommand \Eprint [0]{\href }%
\providecommand \doibase [0]{http://dx.doi.org/}%
\providecommand \selectlanguage [0]{\@gobble}%
\providecommand \bibinfo  [0]{\@secondoftwo}%
\providecommand \bibfield  [0]{\@secondoftwo}%
\providecommand \translation [1]{[#1]}%
\providecommand \BibitemOpen [0]{}%
\providecommand \bibitemStop [0]{}%
\providecommand \bibitemNoStop [0]{.\EOS\space}%
\providecommand \EOS [0]{\spacefactor3000\relax}%
\providecommand \BibitemShut  [1]{\csname bibitem#1\endcsname}%
\let\auto@bib@innerbib\@empty
\bibitem [{\citenamefont {Jerrum}\ and\ \citenamefont
  {Sinclair}(1993)}]{jerrum1993polynomial}%
  \BibitemOpen
  \bibfield  {author} {\bibinfo {author} {\bibfnamefont {M.}~\bibnamefont
  {Jerrum}}\ and\ \bibinfo {author} {\bibfnamefont {A.}~\bibnamefont
  {Sinclair}},\ }\href@noop {} {\bibfield  {journal} {\bibinfo  {journal} {SIAM
  Journal on computing}\ }\textbf {\bibinfo {volume} {22}},\ \bibinfo {pages}
  {1087} (\bibinfo {year} {1993})}\BibitemShut {NoStop}%
\bibitem [{\citenamefont {Guo}\ and\ \citenamefont
  {Jerrum}(2016)}]{guo2016random}%
  \BibitemOpen
  \bibfield  {author} {\bibinfo {author} {\bibfnamefont {H.}~\bibnamefont
  {Guo}}\ and\ \bibinfo {author} {\bibfnamefont {M.}~\bibnamefont {Jerrum}},\
  }\href@noop {} {\enquote {\bibinfo {title} {Random cluster dynamics for the
  ising model is rapidly mixing},}\ } (\bibinfo {year} {2016}),\ \Eprint
  {http://arxiv.org/abs/1605.00139} {arXiv:1605.00139 [cs.DS]} \BibitemShut
  {NoStop}%
\bibitem [{\citenamefont {Collevecchio}\ \emph {et~al.}(2016)\citenamefont
  {Collevecchio}, \citenamefont {Garoni}, \citenamefont {Hyndman},\ and\
  \citenamefont {Tokarev}}]{Collevecchio_2016}%
  \BibitemOpen
  \bibfield  {author} {\bibinfo {author} {\bibfnamefont {A.}~\bibnamefont
  {Collevecchio}}, \bibinfo {author} {\bibfnamefont {T.~M.}\ \bibnamefont
  {Garoni}}, \bibinfo {author} {\bibfnamefont {T.}~\bibnamefont {Hyndman}}, \
  and\ \bibinfo {author} {\bibfnamefont {D.}~\bibnamefont {Tokarev}},\ }\href
  {\doibase 10.1007/s10955-016-1572-2} {\bibfield  {journal} {\bibinfo
  {journal} {Journal of Statistical Physics}\ }\textbf {\bibinfo {volume}
  {164}},\ \bibinfo {pages} {1082–1102} (\bibinfo {year} {2016})}\BibitemShut
  {NoStop}%
\bibitem [{\citenamefont {Aarts}\ and\ \citenamefont
  {Stamatescu}(2008)}]{Aarts:2008rr}%
  \BibitemOpen
  \bibfield  {author} {\bibinfo {author} {\bibfnamefont {G.}~\bibnamefont
  {Aarts}}\ and\ \bibinfo {author} {\bibfnamefont {I.-O.}\ \bibnamefont
  {Stamatescu}},\ }\href {\doibase 10.1088/1126-6708/2008/09/018} {\bibfield
  {journal} {\bibinfo  {journal} {JHEP}\ }\textbf {\bibinfo {volume} {09}},\
  \bibinfo {pages} {018} (\bibinfo {year} {2008})},\ \Eprint
  {http://arxiv.org/abs/0807.1597} {arXiv:0807.1597 [hep-lat]} \BibitemShut
  {NoStop}%
\bibitem [{\citenamefont {Langfeld}\ and\ \citenamefont
  {Lucini}(2016)}]{Langfeld:2016mct}%
  \BibitemOpen
  \bibfield  {author} {\bibinfo {author} {\bibfnamefont {K.}~\bibnamefont
  {Langfeld}}\ and\ \bibinfo {author} {\bibfnamefont {B.}~\bibnamefont
  {Lucini}},\ }\bibfield  {booktitle} {\emph {\bibinfo {booktitle}
  {{Proceedings, International Meeting Excited QCD 2016: Costa da Caparica,
  Portugal, March 6-12, 2016}}},\ }\href {\doibase 10.5506/APhysPolBSupp.9.503}
  {\bibfield  {journal} {\bibinfo  {journal} {Acta Phys. Polon. Supp.}\
  }\textbf {\bibinfo {volume} {9}},\ \bibinfo {pages} {503} (\bibinfo {year}
  {2016})},\ \Eprint {http://arxiv.org/abs/1606.03879} {arXiv:1606.03879
  [hep-lat]} \BibitemShut {NoStop}%
\bibitem [{\citenamefont {Alexandru}\ \emph {et~al.}(2005)\citenamefont
  {Alexandru}, \citenamefont {Faber}, \citenamefont {Horvath},\ and\
  \citenamefont {Liu}}]{Alexandru:2005ix}%
  \BibitemOpen
  \bibfield  {author} {\bibinfo {author} {\bibfnamefont {A.}~\bibnamefont
  {Alexandru}}, \bibinfo {author} {\bibfnamefont {M.}~\bibnamefont {Faber}},
  \bibinfo {author} {\bibfnamefont {I.}~\bibnamefont {Horvath}}, \ and\
  \bibinfo {author} {\bibfnamefont {K.-F.}\ \bibnamefont {Liu}},\ }\href
  {\doibase 10.1103/PhysRevD.72.114513} {\bibfield  {journal} {\bibinfo
  {journal} {Phys. Rev.}\ }\textbf {\bibinfo {volume} {D72}},\ \bibinfo {pages}
  {114513} (\bibinfo {year} {2005})},\ \Eprint
  {http://arxiv.org/abs/hep-lat/0507020} {arXiv:hep-lat/0507020 [hep-lat]}
  \BibitemShut {NoStop}%
\bibitem [{\citenamefont {de~Forcrand}\ and\ \citenamefont
  {Kratochvila}(2006)}]{deForcrand:2006ec}%
  \BibitemOpen
  \bibfield  {author} {\bibinfo {author} {\bibfnamefont {P.}~\bibnamefont
  {de~Forcrand}}\ and\ \bibinfo {author} {\bibfnamefont {S.}~\bibnamefont
  {Kratochvila}},\ }\bibfield  {booktitle} {\emph {\bibinfo {booktitle}
  {{Hadron physics, proceedings of the Workshop on Computational Hadron
  Physics, University of Cyprus, Nicosia, Cyprus, 14-17 September 2005}}},\
  }\href {\doibase 10.1016/j.nuclphysbps.2006.01.007} {\bibfield  {journal}
  {\bibinfo  {journal} {Nucl. Phys. Proc. Suppl.}\ }\textbf {\bibinfo {volume}
  {153}},\ \bibinfo {pages} {62} (\bibinfo {year} {2006})},\ \bibinfo {note}
  {[,62(2006)]},\ \Eprint {http://arxiv.org/abs/hep-lat/0602024}
  {arXiv:hep-lat/0602024 [hep-lat]} \BibitemShut {NoStop}%
\bibitem [{\citenamefont {Fodor}\ and\ \citenamefont
  {Katz}(2002)}]{Fodor:2001au}%
  \BibitemOpen
  \bibfield  {author} {\bibinfo {author} {\bibfnamefont {Z.}~\bibnamefont
  {Fodor}}\ and\ \bibinfo {author} {\bibfnamefont {S.~D.}\ \bibnamefont
  {Katz}},\ }\href {\doibase 10.1016/S0370-2693(02)01583-6} {\bibfield
  {journal} {\bibinfo  {journal} {Phys. Lett.}\ }\textbf {\bibinfo {volume}
  {B534}},\ \bibinfo {pages} {87} (\bibinfo {year} {2002})},\ \Eprint
  {http://arxiv.org/abs/hep-lat/0104001} {arXiv:hep-lat/0104001 [hep-lat]}
  \BibitemShut {NoStop}%
\bibitem [{\citenamefont {Allton}\ \emph {et~al.}(2002)\citenamefont {Allton},
  \citenamefont {Ejiri}, \citenamefont {Hands}, \citenamefont {Kaczmarek},
  \citenamefont {Karsch}, \citenamefont {Laermann}, \citenamefont {Schmidt},\
  and\ \citenamefont {Scorzato}}]{Allton:2002zi}%
  \BibitemOpen
  \bibfield  {author} {\bibinfo {author} {\bibfnamefont {C.~R.}\ \bibnamefont
  {Allton}}, \bibinfo {author} {\bibfnamefont {S.}~\bibnamefont {Ejiri}},
  \bibinfo {author} {\bibfnamefont {S.~J.}\ \bibnamefont {Hands}}, \bibinfo
  {author} {\bibfnamefont {O.}~\bibnamefont {Kaczmarek}}, \bibinfo {author}
  {\bibfnamefont {F.}~\bibnamefont {Karsch}}, \bibinfo {author} {\bibfnamefont
  {E.}~\bibnamefont {Laermann}}, \bibinfo {author} {\bibfnamefont
  {C.}~\bibnamefont {Schmidt}}, \ and\ \bibinfo {author} {\bibfnamefont
  {L.}~\bibnamefont {Scorzato}},\ }\href {\doibase 10.1103/PhysRevD.66.074507}
  {\bibfield  {journal} {\bibinfo  {journal} {Phys. Rev.}\ }\textbf {\bibinfo
  {volume} {D66}},\ \bibinfo {pages} {074507} (\bibinfo {year} {2002})},\
  \Eprint {http://arxiv.org/abs/hep-lat/0204010} {arXiv:hep-lat/0204010
  [hep-lat]} \BibitemShut {NoStop}%
\bibitem [{\citenamefont {Chandrasekharan}(2013)}]{Chandrasekharan:2013rpa}%
  \BibitemOpen
  \bibfield  {author} {\bibinfo {author} {\bibfnamefont {S.}~\bibnamefont
  {Chandrasekharan}},\ }\href {\doibase 10.1140/epja/i2013-13090-y} {\bibfield
  {journal} {\bibinfo  {journal} {Eur. Phys. J.}\ }\textbf {\bibinfo {volume}
  {A49}},\ \bibinfo {pages} {90} (\bibinfo {year} {2013})},\ \Eprint
  {http://arxiv.org/abs/1304.4900} {arXiv:1304.4900 [hep-lat]} \BibitemShut
  {NoStop}%
\bibitem [{\citenamefont {Alexandru}\ \emph {et~al.}(2020)\citenamefont
  {Alexandru}, \citenamefont {Basar}, \citenamefont {Bedaque},\ and\
  \citenamefont {Warrington}}]{Alexandru:2020wrj}%
  \BibitemOpen
  \bibfield  {author} {\bibinfo {author} {\bibfnamefont {A.}~\bibnamefont
  {Alexandru}}, \bibinfo {author} {\bibfnamefont {G.}~\bibnamefont {Basar}},
  \bibinfo {author} {\bibfnamefont {P.~F.}\ \bibnamefont {Bedaque}}, \ and\
  \bibinfo {author} {\bibfnamefont {N.~C.}\ \bibnamefont {Warrington}},\
  }\href@noop {} {\  (\bibinfo {year} {2020})},\ \Eprint
  {http://arxiv.org/abs/2007.05436} {arXiv:2007.05436 [hep-lat]} \BibitemShut
  {NoStop}%
\bibitem [{\citenamefont {de~Forcrand}\ and\ \citenamefont
  {Philipsen}(2007)}]{deForcrand:2006pv}%
  \BibitemOpen
  \bibfield  {author} {\bibinfo {author} {\bibfnamefont {P.}~\bibnamefont
  {de~Forcrand}}\ and\ \bibinfo {author} {\bibfnamefont {O.}~\bibnamefont
  {Philipsen}},\ }\href {\doibase 10.1088/1126-6708/2007/01/077} {\bibfield
  {journal} {\bibinfo  {journal} {JHEP}\ }\textbf {\bibinfo {volume} {01}},\
  \bibinfo {pages} {077} (\bibinfo {year} {2007})},\ \Eprint
  {http://arxiv.org/abs/hep-lat/0607017} {arXiv:hep-lat/0607017 [hep-lat]}
  \BibitemShut {NoStop}%
\bibitem [{\citenamefont {Thirring}(1958)}]{thirring1958soluble}%
  \BibitemOpen
  \bibfield  {author} {\bibinfo {author} {\bibfnamefont {W.~E.}\ \bibnamefont
  {Thirring}},\ }\href@noop {} {\bibfield  {journal} {\bibinfo  {journal}
  {Annals of Physics}\ }\textbf {\bibinfo {volume} {3}},\ \bibinfo {pages} {91}
  (\bibinfo {year} {1958})}\BibitemShut {NoStop}%
\bibitem [{\citenamefont {Alexandru}\ \emph {et~al.}(2017)\citenamefont
  {Alexandru}, \citenamefont {Basar}, \citenamefont {Bedaque}, \citenamefont
  {Ridgway},\ and\ \citenamefont {Warrington}}]{Alexandru:2016ejd}%
  \BibitemOpen
  \bibfield  {author} {\bibinfo {author} {\bibfnamefont {A.}~\bibnamefont
  {Alexandru}}, \bibinfo {author} {\bibfnamefont {G.}~\bibnamefont {Basar}},
  \bibinfo {author} {\bibfnamefont {P.~F.}\ \bibnamefont {Bedaque}}, \bibinfo
  {author} {\bibfnamefont {G.~W.}\ \bibnamefont {Ridgway}}, \ and\ \bibinfo
  {author} {\bibfnamefont {N.~C.}\ \bibnamefont {Warrington}},\ }\href
  {\doibase 10.1103/PhysRevD.95.014502} {\bibfield  {journal} {\bibinfo
  {journal} {Phys. Rev. D}\ }\textbf {\bibinfo {volume} {95}},\ \bibinfo
  {pages} {014502} (\bibinfo {year} {2017})},\ \Eprint
  {http://arxiv.org/abs/1609.01730} {arXiv:1609.01730 [hep-lat]} \BibitemShut
  {NoStop}%
\bibitem [{\citenamefont {Alexandru}\ \emph
  {et~al.}(2016{\natexlab{a}})\citenamefont {Alexandru}, \citenamefont {Basar},
  \citenamefont {Bedaque}, \citenamefont {Ridgway},\ and\ \citenamefont
  {Warrington}}]{Alexandru:2015sua}%
  \BibitemOpen
  \bibfield  {author} {\bibinfo {author} {\bibfnamefont {A.}~\bibnamefont
  {Alexandru}}, \bibinfo {author} {\bibfnamefont {G.}~\bibnamefont {Basar}},
  \bibinfo {author} {\bibfnamefont {P.~F.}\ \bibnamefont {Bedaque}}, \bibinfo
  {author} {\bibfnamefont {G.~W.}\ \bibnamefont {Ridgway}}, \ and\ \bibinfo
  {author} {\bibfnamefont {N.~C.}\ \bibnamefont {Warrington}},\ }\href
  {\doibase 10.1007/JHEP05(2016)053} {\bibfield  {journal} {\bibinfo  {journal}
  {JHEP}\ }\textbf {\bibinfo {volume} {05}},\ \bibinfo {pages} {053} (\bibinfo
  {year} {2016}{\natexlab{a}})},\ \Eprint {http://arxiv.org/abs/1512.08764}
  {arXiv:1512.08764 [hep-lat]} \BibitemShut {NoStop}%
\bibitem [{\citenamefont {Pawlowski}\ and\ \citenamefont
  {Zielinski}(2013)}]{Pawlowski:2013gag}%
  \BibitemOpen
  \bibfield  {author} {\bibinfo {author} {\bibfnamefont {J.~M.}\ \bibnamefont
  {Pawlowski}}\ and\ \bibinfo {author} {\bibfnamefont {C.}~\bibnamefont
  {Zielinski}},\ }\href {\doibase 10.1103/PhysRevD.87.094509} {\bibfield
  {journal} {\bibinfo  {journal} {Phys. Rev. D}\ }\textbf {\bibinfo {volume}
  {87}},\ \bibinfo {pages} {094509} (\bibinfo {year} {2013})},\ \Eprint
  {http://arxiv.org/abs/1302.2249} {arXiv:1302.2249 [hep-lat]} \BibitemShut
  {NoStop}%
\bibitem [{\citenamefont {Alexandru}\ \emph
  {et~al.}(2016{\natexlab{b}})\citenamefont {Alexandru}, \citenamefont
  {Basar},\ and\ \citenamefont {Bedaque}}]{Alexandru:2015xva}%
  \BibitemOpen
  \bibfield  {author} {\bibinfo {author} {\bibfnamefont {A.}~\bibnamefont
  {Alexandru}}, \bibinfo {author} {\bibfnamefont {G.}~\bibnamefont {Basar}}, \
  and\ \bibinfo {author} {\bibfnamefont {P.}~\bibnamefont {Bedaque}},\ }\href
  {\doibase 10.1103/PhysRevD.93.014504} {\bibfield  {journal} {\bibinfo
  {journal} {Phys. Rev. D}\ }\textbf {\bibinfo {volume} {93}},\ \bibinfo
  {pages} {014504} (\bibinfo {year} {2016}{\natexlab{b}})},\ \Eprint
  {http://arxiv.org/abs/1510.03258} {arXiv:1510.03258 [hep-lat]} \BibitemShut
  {NoStop}%
\bibitem [{\citenamefont {Montvay}\ and\ \citenamefont
  {M{\"u}nster}(1997)}]{montvay1997quantum}%
  \BibitemOpen
  \bibfield  {author} {\bibinfo {author} {\bibfnamefont {I.}~\bibnamefont
  {Montvay}}\ and\ \bibinfo {author} {\bibfnamefont {G.}~\bibnamefont
  {M{\"u}nster}},\ }\href@noop {} {\emph {\bibinfo {title} {Quantum fields on a
  lattice}}}\ (\bibinfo  {publisher} {Cambridge University Press},\ \bibinfo
  {year} {1997})\BibitemShut {NoStop}%
\bibitem [{\citenamefont {Pawlowski}\ \emph {et~al.}(2015)\citenamefont
  {Pawlowski}, \citenamefont {Stamatescu},\ and\ \citenamefont
  {Zielinski}}]{Pawlowski_2015}%
  \BibitemOpen
  \bibfield  {author} {\bibinfo {author} {\bibfnamefont {J.~M.}\ \bibnamefont
  {Pawlowski}}, \bibinfo {author} {\bibfnamefont {I.-O.}\ \bibnamefont
  {Stamatescu}}, \ and\ \bibinfo {author} {\bibfnamefont {C.}~\bibnamefont
  {Zielinski}},\ }\href {\doibase 10.1103/physrevd.92.014508} {\bibfield
  {journal} {\bibinfo  {journal} {Physical Review D}\ }\textbf {\bibinfo
  {volume} {92}} (\bibinfo {year} {2015}),\
  10.1103/physrevd.92.014508}\BibitemShut {NoStop}%
\bibitem [{\citenamefont {Alexandru}\ \emph
  {et~al.}(2018{\natexlab{a}})\citenamefont {Alexandru}, \citenamefont
  {Bedaque}, \citenamefont {Lamm},\ and\ \citenamefont
  {Lawrence}}]{Alexandru:2018fqp}%
  \BibitemOpen
  \bibfield  {author} {\bibinfo {author} {\bibfnamefont {A.}~\bibnamefont
  {Alexandru}}, \bibinfo {author} {\bibfnamefont {P.~F.}\ \bibnamefont
  {Bedaque}}, \bibinfo {author} {\bibfnamefont {H.}~\bibnamefont {Lamm}}, \
  and\ \bibinfo {author} {\bibfnamefont {S.}~\bibnamefont {Lawrence}},\ }\href
  {\doibase 10.1103/PhysRevD.97.094510} {\bibfield  {journal} {\bibinfo
  {journal} {Phys. Rev.}\ }\textbf {\bibinfo {volume} {D97}},\ \bibinfo {pages}
  {094510} (\bibinfo {year} {2018}{\natexlab{a}})},\ \Eprint
  {http://arxiv.org/abs/1804.00697} {arXiv:1804.00697 [hep-lat]} \BibitemShut
  {NoStop}%
\bibitem [{\citenamefont {Alexandru}\ \emph
  {et~al.}(2018{\natexlab{b}})\citenamefont {Alexandru}, \citenamefont
  {Bedaque}, \citenamefont {Lamm}, \citenamefont {Lawrence},\ and\
  \citenamefont {Warrington}}]{Alexandru:2018ddf}%
  \BibitemOpen
  \bibfield  {author} {\bibinfo {author} {\bibfnamefont {A.}~\bibnamefont
  {Alexandru}}, \bibinfo {author} {\bibfnamefont {P.~F.}\ \bibnamefont
  {Bedaque}}, \bibinfo {author} {\bibfnamefont {H.}~\bibnamefont {Lamm}},
  \bibinfo {author} {\bibfnamefont {S.}~\bibnamefont {Lawrence}}, \ and\
  \bibinfo {author} {\bibfnamefont {N.~C.}\ \bibnamefont {Warrington}},\ }\href
  {\doibase 10.1103/PhysRevLett.121.191602} {\bibfield  {journal} {\bibinfo
  {journal} {Phys. Rev. Lett.}\ }\textbf {\bibinfo {volume} {121}},\ \bibinfo
  {pages} {191602} (\bibinfo {year} {2018}{\natexlab{b}})},\ \Eprint
  {http://arxiv.org/abs/1808.09799} {arXiv:1808.09799 [hep-lat]} \BibitemShut
  {NoStop}%
\bibitem [{\citenamefont {Kashiwa}\ \emph {et~al.}(2019)\citenamefont
  {Kashiwa}, \citenamefont {Mori},\ and\ \citenamefont
  {Ohnishi}}]{Kashiwa:2019lkv}%
  \BibitemOpen
  \bibfield  {author} {\bibinfo {author} {\bibfnamefont {K.}~\bibnamefont
  {Kashiwa}}, \bibinfo {author} {\bibfnamefont {Y.}~\bibnamefont {Mori}}, \
  and\ \bibinfo {author} {\bibfnamefont {A.}~\bibnamefont {Ohnishi}},\ }\href
  {\doibase 10.1103/PhysRevD.99.114005} {\bibfield  {journal} {\bibinfo
  {journal} {Phys. Rev. D}\ }\textbf {\bibinfo {volume} {99}},\ \bibinfo
  {pages} {114005} (\bibinfo {year} {2019})},\ \Eprint
  {http://arxiv.org/abs/1903.03679} {arXiv:1903.03679 [hep-lat]} \BibitemShut
  {NoStop}%
\bibitem [{\citenamefont {Ohnishi}\ \emph {et~al.}(2019)\citenamefont
  {Ohnishi}, \citenamefont {Mori},\ and\ \citenamefont
  {Kashiwa}}]{Ohnishi:2019ljc}%
  \BibitemOpen
  \bibfield  {author} {\bibinfo {author} {\bibfnamefont {A.}~\bibnamefont
  {Ohnishi}}, \bibinfo {author} {\bibfnamefont {Y.}~\bibnamefont {Mori}}, \
  and\ \bibinfo {author} {\bibfnamefont {K.}~\bibnamefont {Kashiwa}},\ }\href
  {\doibase 10.7566/JPSCP.26.024011} {\bibfield  {journal} {\bibinfo  {journal}
  {JPS Conf. Proc.}\ }\textbf {\bibinfo {volume} {26}},\ \bibinfo {pages}
  {024011} (\bibinfo {year} {2019})}\BibitemShut {NoStop}%
\bibitem [{\citenamefont {Lawrence}(2020)}]{Lawrence:2020irw}%
  \BibitemOpen
  \bibfield  {author} {\bibinfo {author} {\bibfnamefont {S.}~\bibnamefont
  {Lawrence}},\ }\emph {\bibinfo {title} {{Sign Problems in Quantum Field
  Theory: Classical and Quantum Approaches}}},\ \href@noop {} {Ph.D. thesis}
  (\bibinfo {year} {2020}),\ \Eprint {http://arxiv.org/abs/2006.03683}
  {arXiv:2006.03683 [hep-lat]} \BibitemShut {NoStop}%
\bibitem [{\citenamefont {Lawrence}(2018)}]{Lawrence:2018mve}%
  \BibitemOpen
  \bibfield  {author} {\bibinfo {author} {\bibfnamefont {S.}~\bibnamefont
  {Lawrence}},\ }\bibfield  {booktitle} {\emph {\bibinfo {booktitle}
  {{Proceedings, 36th International Symposium on Lattice Field Theory (Lattice
  2018): East Lansing, MI, United States, July 22-28, 2018}}},\ }\href
  {\doibase 10.22323/1.334.0149} {\bibfield  {journal} {\bibinfo  {journal}
  {PoS}\ }\textbf {\bibinfo {volume} {LATTICE2018}},\ \bibinfo {pages} {149}
  (\bibinfo {year} {2018})},\ \Eprint {http://arxiv.org/abs/1810.06529}
  {arXiv:1810.06529 [hep-lat]} \BibitemShut {NoStop}%
\bibitem [{\citenamefont {Alexandru}\ \emph
  {et~al.}(2016{\natexlab{c}})\citenamefont {Alexandru}, \citenamefont {Basar},
  \citenamefont {Bedaque}, \citenamefont {Vartak},\ and\ \citenamefont
  {Warrington}}]{Alexandru:2016gsd}%
  \BibitemOpen
  \bibfield  {author} {\bibinfo {author} {\bibfnamefont {A.}~\bibnamefont
  {Alexandru}}, \bibinfo {author} {\bibfnamefont {G.}~\bibnamefont {Basar}},
  \bibinfo {author} {\bibfnamefont {P.~F.}\ \bibnamefont {Bedaque}}, \bibinfo
  {author} {\bibfnamefont {S.}~\bibnamefont {Vartak}}, \ and\ \bibinfo {author}
  {\bibfnamefont {N.~C.}\ \bibnamefont {Warrington}},\ }\href {\doibase
  10.1103/PhysRevLett.117.081602} {\bibfield  {journal} {\bibinfo  {journal}
  {Phys. Rev. Lett.}\ }\textbf {\bibinfo {volume} {117}},\ \bibinfo {pages}
  {081602} (\bibinfo {year} {2016}{\natexlab{c}})},\ \Eprint
  {http://arxiv.org/abs/1605.08040} {arXiv:1605.08040 [hep-lat]} \BibitemShut
  {NoStop}%
\bibitem [{\citenamefont {Detmold}\ \emph {et~al.}(2020)\citenamefont
  {Detmold}, \citenamefont {Kanwar}, \citenamefont {Wagman},\ and\
  \citenamefont {Warrington}}]{Detmold:2020ncp}%
  \BibitemOpen
  \bibfield  {author} {\bibinfo {author} {\bibfnamefont {W.}~\bibnamefont
  {Detmold}}, \bibinfo {author} {\bibfnamefont {G.}~\bibnamefont {Kanwar}},
  \bibinfo {author} {\bibfnamefont {M.~L.}\ \bibnamefont {Wagman}}, \ and\
  \bibinfo {author} {\bibfnamefont {N.~C.}\ \bibnamefont {Warrington}},\ }\href
  {\doibase 10.1103/PhysRevD.102.014514} {\bibfield  {journal} {\bibinfo
  {journal} {Phys. Rev. D}\ }\textbf {\bibinfo {volume} {102}},\ \bibinfo
  {pages} {014514} (\bibinfo {year} {2020})},\ \Eprint
  {http://arxiv.org/abs/2003.05914} {arXiv:2003.05914 [hep-lat]} \BibitemShut
  {NoStop}%
\end{thebibliography}%
\end{document}